\documentstyle[prl,aps,preprint]{revtex}
\newcommand{\be}{\begin{equation}}
\newcommand{\ee}{\end{equation}}
\newcommand{\ba}{\begin{eqnarray}}
\newcommand{\ea}{\end{eqnarray}}
\newcommand{\bec}{\begin{center}}
\newcommand{\eec}{\end{center}}

\let\ssection=\section
\renewcommand{\section}{\setcounter{equation}{0}\ssection}

\begin{document}
\draft

%\twocolumn
\widetext

\title{Pleba\'nski--Demia\'nski--like solutions in metric--affine gravity}

\author{
Alberto Garc\'{\i}a$^\star$\thanks{E-mail: aagarcia@fis.cinvestav.mx}, 
Friedrich W. Hehl$^{\#}$\thanks{E-mail: hehl@thp.uni-koeln.de},
Claus L\"ammerzahl$^\triangleleft$\thanks{E-mail: 
claus@spock.physik.uni-konstanz.de}, \\
Alfredo Mac\'{\i}as$^\diamond$\thanks{E-mail: amac@xanum.uam.mx},
and Jos\'e Socorro$^{\$}$\thanks{E-mail: socorro@ifug4.ugto.mx}\\
$^{\star}$  Departamento de F\'{\i}sica,\\
CINVESTAV--IPN, Apartado Postal 14--740, C.P. 07000, M\'exico,  
D.F., Mexico\\
$^{\#}$Institute for Theoretical Physics,\\
 University of Cologne, D--50923 K\"oln, Germany\\
$^\triangleleft$ Fakult\"at f\"ur Physik, Universit\"at Konstanz \\ 
Postfach 5560 M674, D--78434 Konstanz, Germany \\
$^{\diamond}$ Departamento de F\'{\i}sica,\\
Universidad Aut\'onoma Metropolitana--Iztapalapa,\\
Apartado Postal 55-534, C.P. 09340, M\'exico, D.F., Mexico.\\
$^{\$}$ Instituto de F\'{\i}sica de la Universidad de Guanajuato,\\
Apartado Postal E-143, C.P. 37150, Le\'on, Guanajuato, Mexico.
}

\date{\today}

\maketitle

\begin{abstract}
  We consider a (non--Riemannian) metric--affine gravity theory, in
  particular its nonmetricity--torsion sector ``isomorphic'' to the
  Einstein--Maxwell theory. We map certain Einstein--Maxwell
  electrovacuum solutions to it, namely the Pleba\'nski--Demia\'nski
  class of Petrov type D metrics. {\em file plebdem3.tex, 1998-02-27}

\end{abstract}
\vspace{0.5cm}


\narrowtext
%**********************************************************
\section{Introduction}

Nowadays, there exists a revival of interest in metric--affine gravity
(MAG) theories. It has been demonstrated that they contain the
axi--dilatonic sector of {\em low energy string theory} \cite{dt95} as
special case. Moreover, the gravitational interactions involving the
axion and dilaton may be derived from a geometrical action principle
involving the curvature scalar with a non--Riemannian connection. In
other words, the axi--dilatonic sector of the low energy string theory
can be expressed in terms of a geometry with {\em torsion} and {\em
  nonmetricity} \cite{dot95}. This formulation emphasizes the
geometrical nature of the axion and dilaton fields and raises
questions about the most appropriate geometry for the discussion of
physical phenomena involving these fields.
 
Recently, it has been proposed that certain MAG models can be reduced
to an {\em effective} Einstein--Proca system \cite{De96,heob97}.
Indeed, we have in these kind of models, beside the orthonormal
coframe of spacetime, effectively only one extra one--form (co-vector)
field as additional degree of freedom.

Very important classes of Petrov type D solutions of the
Einstein--Maxwell equations are the Pleba\'nski classes. The most
general of them is the so--called Pleba\'nski--Demia\'nski solution
\cite{pd}, which, as is well known, contains as special cases, among
others, the Pleba\'nski--Carter, the Kerr--Newman, and the Kerr
solutions \cite{mg}.  In this paper we are going to map this complete
space of general electrovacuum solutions to a metric--affine gravity
model, which generalizes Einstein's general relativity.  Thus, we are
able to present solutions to this MAG model and to give to these
solutions a physical interpretation.

One arrives at the metric--affine gauge theory of gravity if one
gauges the affine group and additionally allows for a metric $g$
\cite{PR}.  The four--dimensional affine group $A(4,R)$ is the
semidirect product of the {\it translation} group $R^4$ and the {\it
  linear group} $GL(4,R)$, that is, $GL(4,R)=R^4
{\;{\rlap{$\supset$}\times}\;} GL(4\,,R)$. The spacetime of MAG
encompasses two different post--Riemannian structures: the
nonmetricity one--form $Q_{\alpha\beta}=Q_{i\alpha\beta} \,dx^i$ and
the torsion two--form $T^\alpha =\frac{1}{2}\,T_{ij}{}^\alpha
dx^i\wedge dx^j$.  According to the Yang--Mills fashion, gauge
Lagrangians of MAG are quadratic in curvature, torsion, and
nonmetricity.  One way to investigate the potentialities of such
models is to look for {\em exact} solutions.

The search for exact solutions of MAG began with the work of
Tres\-guerres\cite{Tres14,Tres15}, Tucker and Wang\cite{TW}, Obukhov
et al. \cite{heh96}, Vlashinsky et al. \cite{he96}, and of Puntigam et
al. \cite{PLH97}.  Mac\'{\i}as et al. \cite{ma}, and Socorro et al.
\cite{slmm} mapped the Einstein--Maxwell sector of dilaton--gravity,
emerging from low energy string theory, and found new soliton and
multipole solutions of MAG.  However, it is important to note that in
order to incorporate the scalar dilaton field, one could, for
instance, generalize the torsion kink of Baekler et al. \cite{kink},
an exact solution with an external massless scalar field, or one could
turn to the axi--dilatonic sector of MAG \cite{dt95}.  Moreover,
solutions implying the existence of torsion shock waves have already
been found by Garc\'{\i}a et al. \cite{glmms}.  In this spirit, we are
going to look for a wide class of solutions of the vacuum field
equations of MAG.  Note that such a solution with the additional
electromagnetic field of a point charge has been presented in
\cite{PLH97}.  There it was confirmed that the electromagnetic field
is not directly influenced by the post--Riemannian structures torsion
and nonmetricity.

A general quadratic Lagrangian in MAG reads \cite{heob97,PR}:
\begin{eqnarray} 
\label{QMA} V_{\rm MAG}&=&
\frac{1}{2\kappa}\,\left[-a_0\,R^{\alpha\beta}\wedge\eta_{\alpha\beta} 
-2\lambda_{\hbox{\scriptsize cosm}}\,\eta+  
T^\alpha\wedge{}^*\!\left(\sum_{I=1}^{3}a_{I}\,^{(I)}
T_\alpha\right)\right.\nonumber\\
& & + \left.  2\left(\sum_{I=2}^{4}c_{I}\,^{(I)}Q_{\alpha\beta}\right)
\wedge\vartheta^\alpha\wedge{}^*\!\, T^\beta + Q_{\alpha\beta}
\wedge{}^*\!\left(\sum_{I=1}^{4}b_{I}\,^{(I)}Q^{\alpha\beta}\right)\right] 
\nonumber\\ 
& & - \frac{1}{2}\,R^{\alpha\beta} \wedge{}^*\!
\left(\sum_{I=1}^{6}w_{I}\,^{(I)}W_{\alpha\beta} +
  \sum_{I=1}^{5}{z}_{I}\,^{(I)}Z_{\alpha\beta}\right)
\label{lobo}\,.
\end{eqnarray} 
The signature of spacetime is $(- + + +)$, the volume four--form
$\eta:={}^*\!\, 1$, the two--form
$\eta_{\alpha\beta}:=\,^*(\vartheta_\alpha\wedge\vartheta_\beta)$, and
the dimensionless coupling constants read
\begin{equation} \label{constants}
  a_0, \ldots a_3, \, b_1, \ldots b_4, \, c_2, c_3,c_4, \, w_1, \ldots
  w_6, \, z_1, \ldots z_5
\label{consts}\,.
\end{equation} 
Moreover, $\kappa$ is the gravitational and
$\lambda_{\hbox{\scriptsize cosm}}$ the cosmological constant. In
suitable units, $\kappa=1$, which will be assumed in future. In the
curvature square term we introduced the antisymmetric part
$W_{\alpha\beta}:= R_{[\alpha\beta]}$ and the symmetric part
$Z_{\alpha\beta}:= R_{(\alpha\beta)}$ of the curvature two--form. In
$Z_{\alpha\beta}$, we meet a purely post--Riemannian part. Weyl's
segmental curvature $^{(4)}Z_{\alpha\beta}:=
R_\gamma{}^\gamma\,g_{\alpha\beta}/4= g_{\alpha \beta}\, dQ$, with the
Weyl covector $Q:=Q_\gamma{}^\gamma/4$, has {\em formally} a similar
structure as the electromagnetic field strength $F=dA$, but is
physically quite different since it is related to Weyl rescalings.

For the torsion and nonmetricity field configurations, we concentrate
on the simplest non--trivial case {\em with} shear. According to its
irreducible decomposition \cite{PR}, the nonmetricity contains two
covector pieces, namely $^{(4)}Q_{\alpha\beta}= Q\,g_{\alpha\beta}$,
the dilation piece, and
\begin{equation}
 ^{(3)}Q_{\alpha\beta}={4\over 9}\left(\vartheta_{(\alpha}e_{\beta)}\rfloor 
\Lambda - {1\over 4}
g_{\alpha\beta}\Lambda\right)\,,\qquad \hbox{with}\qquad
  \Lambda:= \vartheta^{\alpha}e^{\beta}\rfloor\!
  {\nearrow\!\!\!\!\!\!\!Q}_{\alpha\beta}\label{3q}\,,
\end{equation}
a proper shear piece. Accordingly, our ansatz for the nonmetricity  
reads
\begin{equation}
  Q_{\alpha\beta}=\, ^{(3)}Q_{\alpha\beta} +\,
  ^{(4)}Q_{\alpha\beta}\,.\label{QQ}
\end{equation}
The torsion, in addition to its tensor piece,
encompasses a covector and an axial covector piece. Let us choose only
the covector piece as non--vanishing:
\begin{equation}
T^{\alpha}={}^{(2)}T^{\alpha}={1\over 3}\,\vartheta^{\alpha}\wedge  
T\,,
\qquad \hbox{with}\qquad T:=e_{\alpha}\rfloor T^{\alpha}\,.\label{TT}
\end{equation}
Thus we are left with the three non--trivial one--forms $Q$, $\Lambda$,
and $T$.  We shall assume that this triplet of one--forms shares the
spacetime symmetries, that is, its members are proportional to each other
\cite{heh96,he96,PLH97,ma}.

With propagating nonmetricity $Q_{\alpha\beta}$ two types of charge
are expected to arise: {\em One dilation charge} related by the
Noether procedure to the trace of the nonmetricity, the Weyl covector
$Q=Q_i dx^i$. It represents the connection associated with gauging the
scale transformations (instead of the $U(1)$--connection in the case
of the Maxwell's field). Furthermore, {\em nine shear charges} are
expected that are related to the remaining traceless piece
${\nearrow\!\!\!\!\!\!\!Q}_{\alpha\beta}:=Q_{\alpha\beta}-Q\,g_{\alpha\beta}$
of the nonmetricity.

The Lagrangian (\ref{QMA}) is very complicated, in particular on
account of its curvature square pieces. Therefore we have to restrict
its generality in order to stay within manageable limits. Our ansatz
for the nonmetricity is expected to require a nonvanishing
post--Riemannian term quadratic in the segmental curvature.
Accordingly, in (\ref{QMA}) we choose
\begin{equation}\label{simplicity}
  w_1=\dots=w_6=0\,,\qquad z_1=z_2=z_3=z_5=0\,,
\end{equation}
that is, only $z_4$ is allowed to survive.

The plan of the paper is as follows: In Sec. 2 a class of solutions,
which is related to the Pleba\'nski--Demia\'nski solution of the
Einstein--Maxwell system, is presented, and in Sec. 3 we shall discuss
the results and the further prospects of the theory.

%************************************************************
\section{Pleba\'nski--Demia\'nski--like solution in MAG}

We start from the coframe of Pleba\'nski and Demia\'nski \cite{pd}
which is specified in terms of the coordinates $(\tau,y,x,\sigma)$:
\begin{eqnarray}\label{coframe} 
\vartheta^{\hat{0}}&=&\frac{1}{H}\sqrt{\frac{Y}
    {\widetilde{\Delta}}}\; (d\tau-x^2d\sigma)\,,\\ 
  \vartheta^{\hat{1}}&=&\frac{1}{H}\sqrt{\frac{\widetilde{\Delta}}{Y}}
  \;dy\,,\\ 
  \vartheta^{\hat{2}}&=&\frac{1}{H}\sqrt{\frac{\widetilde{\Delta}}{X}}
  \;dx\,,\\ 
  \vartheta^{\hat{3}}&=&\frac{1}{H}\sqrt{\frac{X}{\widetilde{\Delta}}}
  \; (d\tau+y^2d\sigma)\,.\label{coframe'}
\end{eqnarray} 
Here $H=H(x,y)$, $X=X(x)$, $Y=Y(y)$, and
$\widetilde{\Delta}=\widetilde{\Delta}(x,y)$ are unknown functions.
The coframe is orthonormal, \be\label{ortho}
g=o_{\alpha\beta}\,\vartheta^\alpha\otimes\vartheta^b\,,\qquad{\rm
  with}\qquad o_{\alpha\beta}={\rm diag}(-1,+1,+1,+1)\,.\ee Thus we
find the following explicit expression for the metric: \be g =
\frac{1}{H^2} \left[ -\frac{Y}{\widetilde{\Delta}} \left( d\tau- x^2
    \, d \sigma\right)^2 + \frac{\widetilde{\Delta}}{Y}\, dy^2 +
  \frac{\widetilde{\Delta}}{X} \,dx^2 + \frac{X}{\widetilde{\Delta}}
  \left( d\tau +y^2 d \sigma \right)^2 \right] \,.  \ee For the
nonmetricity and torsion we assume that they are represented by a {\em
  triplet of one--forms,} the Weyl covector $Q$, the covector $\Lambda$
corresponding to the third irreducible nonmetricity piece, and the
torsion trace $T$.

We substitute the local metric $o_{\alpha\beta}$, the coframe
(\ref{coframe}--\ref{coframe'}), the nonmetricity (\ref{QQ}), and
the torsion (\ref{TT}) into the two field equations following from
the Lagrangian (\ref{lobo}) with (\ref{simplicity}) by variation with
respect to metric and connection. Then, provided the (rather weak)
constraint \ba 32 a_0^2 b_4 - 4a_0 a_2 b_4+64 a_0 b_3 b_4- 32 a_2 b_3
b_4 + 48 a_0 b_4 c_3 + 24 b_4 c_3^2 + 24 b_3 c_4^2 & & \nonumber \\ +
12 a_0 a_2 b_3 +48 a_0 b_3 c_4 - 9a_0 c_3^2 + 18 a_0 c_3 c_4 + 3a_0
c_4^2 +6a_0^2 a_2 + 24 a_0^2 c_4 & = & 0
\label{const} \, , 
\ea on the coupling constants (\ref{consts}) is fulfilled, we find a
general exact solution for the following expressions:
\begin{equation}
  \frac{Q}{k_0}=
  \frac{\Lambda}{k_1}=\frac{T}{k_2}=\frac{H}{\sqrt{\widetilde{\Delta}}}
  \left(\frac{N_{{\rm e}}\,y}{\sqrt{Y}}\;\vartheta^{\hat{0}}+
    \frac{N_{{\rm g}}\,x}{\sqrt{X}}\;\vartheta^{\hat{3}}\right)
\label{sol1}\,,
\end{equation}
\ba H(x,y) &=& 1 - \mu x y \, , \nonumber\\ X(x) &=& (b-g^2) + 2 n x -
\epsilon x^2 +2 m \mu \, x^3 - \left(
  \frac{\lambda_{\hbox{\scriptsize cosm}}}{3a_0}+ \mu^2( b + e^2)
\right)\, x^4 \, , \nonumber\\ Y(y) &=&(b+ e^2) - 2 m y + \epsilon y^2
-2 n \mu \, y^3 - \left( \frac{\lambda_{\hbox{\scriptsize cosm}}}{3a_0}+
  \mu^2( b - g^2) \right)\, y^4 \, , \nonumber\\ 
\widetilde{\Delta}(x,y)&=& x^2 + y^2 \, .
\label{solutions}
\ea Here $N_{\rm e}$ and $N_{\rm g}$ are the quasi--electric and
quasi--magnetic nonmetricity--torsion charges of the source which
fulfill \be \frac{z_4 k_0^2}{2a_0}\left( N_{\rm e}^2+N_{\rm
    g}^2 \right)= g^2+ e^2 \, .
\label{z4}
\ee
The coefficients $k_{0}, k_{1}, k_{2}$ in (\ref{sol1}) are determined
by the dimensionless coupling constants (\ref{consts}) of the
Lagrangian: \ba k_0 &:=& \left({a_2\over 2}-a_0\right)(8b_3 + a_0) -
3(c_3 + a_0 )^2\,,
\label{k0}\\
k_1 &:=& -9\left[ a_0\left({a_2\over 2} - a_0\right) + 
(c_3 + a_0 )(c_4 + a_0 )\right]\,,
\label{k1}\\
k_2 &:=& {3\over 2} \left[ 3a_0 (c_3 + a_0 ) + (
8b_3 + a_0)(c_4 + a_0 )\right]
\label{k2}\, .
\ea
Then the constraint (\ref{const}) can be put into the following more compact 
form
\be
  b_4=\frac{a_0k+2c_4k_2}{8k_0}\,,\qquad\hbox{with}\qquad k:=
  3k_0-k_1+2k_2
\label{b4}\, ,
\ee The constants $\mu$, $b$, $g$, $e$, $n$, and $m$ are free
parameters. The parameter $\epsilon$ is related to the 2--dimensional
spacelike $xy$--surface, it is $\epsilon = 1$ for spherical, $\epsilon
= 0$ for flat, and $\epsilon = -1$ for hyperbolical geometry.

If we collect our results, then the nonmetricity and the torsion read
as follows:
\begin{equation}
  Q^{\alpha\beta} = \left[k_0 \,o^{\alpha\beta} +\frac{4}{9}\, k_1
    \,\left(\vartheta^{(\alpha}e^{\beta )}\rfloor-\frac{1}{4}\,
      o^{\alpha\beta}\right)\right]\;
  \frac{H}{\sqrt{\widetilde{\Delta}}} \left(\frac{N_{\rm e}\,
      y}{\sqrt{Y }}\;\vartheta^{\hat{0}}+\frac{N_{\rm
        g}\,x}{\sqrt{X}}\;\vartheta^{\hat{3}}
    \label{nichtmetrizitaet}\right)\,,
\end{equation}
\begin{equation} 
  T^\alpha = \frac{k_2}{3}\;\vartheta^\alpha\wedge\;
  \frac{H}{\sqrt{\widetilde{\Delta}}} \left(\frac{N_{\rm
        e}\,y}{\sqrt{ Y}} \;\vartheta^{\hat{0}}+\frac{N_{\rm
        g}\,x}{\sqrt{X}} \;\vartheta^{\hat{3}}\right)
  \,.\label{torsion}
\end{equation}
We recognize, see also (\ref{sol1}), that the members $Q,\Lambda,T$ of
the triplet are proportional to each other. Therefore, we have in our
model, besides the spacetime metric, effectively only one extra
one--form as additional degree of freedom.  This makes it clear why a
mapping of our MAG model to the Einstein--Maxwell system and,
accordingly, the use of the Pleba\'nski--Demia\'nski ansatz is
possible, i.e.\ both models have the same number of degrees of
freedom.  Indeed, using this ansatz of Pleba\'nski and Demia\'nski for
a stationary metric and a corresponding ansatz for nonmetricity and
torsion, where we additionally assumed that only co--vector parts of
these post--Riemannian structures are non--vanishing, we arrived at a
general class of solutions for a MAG model.

The physical interpretation of the post--Riemannian parameters of the
solution, as described above, is clear: The dilation (`Weyl') charges
(related to $^{(4)}Q_{\alpha\beta}$) are described by $k_0N_{\rm{e}}$
and $k_0N_{\rm{g}}$, the shear charges (related to
$^{(3)}Q_{\alpha\beta}$) by $k_1N_{\rm{e}}$ and $k_1N_{\rm{g}}$, and,
eventually, the spin charges (related to $^{(2)}T^\alpha$) by
$k_2N_{\rm{e}}$ and $k_2N_{\rm{g}}$, respectively.

The solution (\ref{coframe})--(\ref{ortho}), (\ref{solutions}),
(\ref{nichtmetrizitaet}), and (\ref{torsion}) found above, was checked
with the help of the computer algebra system Reduce
\cite{REDUCE,Stauffer}, using its Excalc package \cite{EXCALC} for
handling exterior differential forms, and by means of the
Reduce--based GRG computer algebra system \cite{GRG}.

%*****************************************************************
\section{Discussion}

The physical motivation to go beyond classical Einstein gravity by
means of MAG models is fairly clear and well founded, see the
discussion in \cite{nehe}. One may suspect that the spin--$3$ modes of
the linear connection in the framework of MAG leads to acausalities.
However, no detailed investigation has been done into this question so
far.  Also, in view of the problems of other theories, like
supergravity and even string {\em field theory} \cite{ne97} in this
respect, it appears unfair to ask questions like that of the
renormalizability of MAG.

Due to the fact that torsion couples to the {\em spin} of matter, a
discussion of those experiments which may lead to restrictions on
torsion also leads, due to (\ref{sol1}), to restrictions on the two
covector parts of the nonmetricity. Therefore, under our triplet
ansatz --- which certainly describes a highly idealized situation ---
it is not necessary to devise separate experiments testing the
coupling of nonmetricity to the shear current of some matter model.

However, spin $\frac{1}{2}$ matter fields couple to the {\em axial}
vector piece $^{(3)}T^\alpha$ of the torsion alone, (massless) gauge
fields carry a helicity of $1$ and do not couple to torsion at all,
see e.g.\ \cite{PLH97}. For massive fields with spin
$s=\frac{1}{2},1,\frac{3}{2}$, we can extract the following formula
for the torsion $T^\alpha$ from the literature, see \cite{Seitz} and
\cite{Spinoza}:
\begin{eqnarray}\label{torsionasseen} T^\alpha_{{\rm as\; seen\; by\; spin\;
      }s>0}&=&\bigl(1-\frac{1}{
    2s}\bigr)\,T^\alpha+\frac{3}{2s}\,^{(3)}T^\alpha\nonumber\\&=&
  \bigl(1-\frac{1}{2s}\bigr)\bigl(^{(1)}T^\alpha+\,^{(2)}T^\alpha\bigr)+
  \bigl(1+\frac{1}{s}\bigr)\,^{(3)}T^\alpha\,. \end{eqnarray} Thereby
we recognize that for massive higher spin fields the trace part
$^{(2)}T^\alpha$ of the torsion couples to the spin of these matter
fields in the same way as the axial part $^{(3)}T^\alpha$, modulo
numerical factors of the order of unity. Accordingly, we can assume
that restrictions on axial torsion also restrict the trace part in a
similar way. Analyzing known experiments, we find, with \cite{Laem},
$t_i\leq 1.5 \times 10^{-15}\,{\hbox{m}}^{-1}$ and, consequently,
$(k_2/k_0)\, Q_i \leq 1.5\times 10^{-15}\,{\hbox{m}}^{-1}$ and
$(k_2/k_1)\,\Lambda_i \leq 1.5\times 10^{-15}\,{\hbox{m}}^{-1}$.  Here
$T=t_i\,dx^i$, $Q=Q_i\,dx^i$, and $\Lambda=\Lambda_i\,dx^i$.

On the other hand, we presented here a complete class of solutions of
MAG. The physical interpretation of the parameters involved in
(\ref{solutions}) can be given as follows: $\mu$ is the acceleration
parameter, $b$ is related to the angular momentum of the solution, $m$
is the mass and $n$ the NUT parameter \cite{mg}. The quasi--electric
and quasi--magnetic charges $e$ and $g$, via (\ref{z4}), are related to
the torsion and nonmetricity charges $N_{\rm e}$ and $N_{\rm g}$,
respectively.

It is important to point out that the generalization of these results
to the whole electrovacuum sector of MAG, i.e., including an
electromagnetic field as source, is straightforward, and these results
will be reported elsewhere.

We want to conclude with two remarks: First, one would like to know at
which energy scale such a MAG framework can be regarded as an {\em
  effective} gravitational model.  According to Ref.\ \cite{PR}, the
motivation for MAG came mainly from particle physics and the manifield
description of an infinite tower of fermions.  One may regard such a
gauge theory of gravity with Weyl invariance as a small but decisive
step towards quantum gravity.  Circumstances under which spacetime
might become non--Riemannian near Planck energies occur in string
theory or in the inflationary model during the early epoch of our
universe.  The simplest such geometry is metric--affine geometry, in
which nonmetricity appears as a field strength, side by side with
torsion and curvature. 

Secondly, on the one hand the axion--dilaton theory emerges at the low
energy limit of string models.  On the other hand such models
represent one sector of the MAG models.  Since these two models have
one important sector in common, we should consider the MAG models, in
a new perspective, as an effective low energy theory of quantum
gravity.

%****************************************

\section{\bf Acknowledgments}

This research was supported by CONACyT, grants No.\ 3544--E9311, No.\
3898P--E9608, and by the joint German--Mexican project Conacyt--DLR
E130--2924 and DLR--Conacyt MXI 6 B0A 6A. Moreover, C.L.\ acknowledges
support from DAAD and Conacyt, and J.S. from the ANUIES--DAAD
agreement, Kennziffer A/98/04459.

\end{document}